\RequirePackage{fix-cm}
\documentclass[twocolumn]{svjour3}          
\smartqed  
\usepackage{graphicx}
\usepackage{amsmath}
\usepackage{amssymb}
\usepackage{color}
\def\simgt{\mathrel{\rlap{\lower 3pt\hbox{$\sim$}} \raise 2.0pt\hbox{$>$}}}

\begin{document}

\title{Gamma Ray Burst studies with THESEUS}


\author{G. Ghirlanda         \and
        R. Salvaterra \and M. Toffano \and S. Ronchini \and C. Guidorzi \and G. Oganesyan \and 
S. Ascenzi \and M.G. Bernardini \and A. E. Camisasca \and S. Mereghetti \and  L. Nava \and M.E. Ravasio \and M. Branchesi \and A. Castro-Tirado \and L. Amati \and A. Blain \and E. Bozzo \and P. O'Brien \and D. G\"{o}tz \and E. Le Floch \and J. P. Osborne \and P. Rosati \and G. Stratta \and N. Tanvir  \and A. I. Bogomazov \and P. D'Avanzo \and M. Hafizi \and S. Mandhai \and A. Melandri  \and A. Peer \and M. Topinka \and S. D. Vergani \and S. Zane 
}

\authorrunning{Ghirlanda, Salvaterra et al.} 

\institute{G. Ghirlanda,  M. G. Bernardini,  L. Nava, P. D'Avanzo, A. Melandri \at
              INAF-Osservatorio Astronomico di Brera, 
              via E. Bianchi 46, 23807 Merate, Italy\\ 
         \email{giancarlo.ghirlanda@inaf.it}       
           \and
           R. Salvaterra, S. Mereghetti, M. Topinka \at
              INAF-Istituto di Astrofisica Spaziale e Fisica cosmica,
              via Alfonso Corti 12, 20133 Milano, Italy\\
	      \email{ruben.salvaterra@inaf.it}
            \and
            S. Ascenzi \at
            Institute of Space Sciences (ICE, CSIC) 
           Campus UAB, Carrer de Can Magrans s/n, 08193, Barcelona, Spain 
            \and
            C. Guidorzi, A. E. Camisasca, P. Rosati \at
            Department of Physics and Earth Science, University of Ferrara, 
           via Saragat 1, 44122 Ferrara, Italy
            \and
           M.E. Ravasio \at
          University of Milano-Bicocca, 
          Piazza della Scienza, 3, 20126 Milan, Italy 
           \and    
           S. Ronchini, G. Oganesyan, M. Branchesi \at
           Gran Sasso Science Institute (GSSI), 
          Viale F. Crispi 7, 67100 L'Aquila, Italy 
           \and
           M. Toffano \at
           University of Insubria, 
          Via Valleggio 11, 22100, Como, Italy 
           \and
          A. Castro-Tirado \at
           Instituto de Astrofísica de Andalucía (IAA-CSIC)
           Glorieta de la Astronomia E18008, Granada, Spain  
           \and
           L. Amati, G. Stratta  \at
          INAF-Osservatorio Astronomico di Bologna 
          via P. Gobetti 101, I40129 Bologna, Italy 
	\and
	A. Blain, P. O'Brien, J. P. Osborne, N. Tanvir, S. Mandhai \at 
	Dept of Physics \& Astronomy,  University of Leicester,  Leicester LE1 7RH, UK 
	\and 
	E. Bozzo \at
	Department of Astronomy, University of Geneva, Ch. d'Ecogia 16, 1290, Versoix (Geneva), Switzerland  
	D. G\"{o}tz , E. Le Floch \at
	AIM-CEA/DRF/Irfu/D\'epartement d’Astrophysique, CNRS, 
	Universit\'e Paris-Saclay, Universit\'e de Paris, 
	Orme des Merisiers, F-91191 Gif-sur-Yvette, France	
	\and
	A. I. Bogomazov \at  
	M. V. Lomonosov Moscow State University, P. K. Sternberg Astronomical Institute, 119234, Universitetkij prospect, 13, Moscow, Russia
	\and 
	M. Hafizi \at
	Department of Physics, Tirana University, Albania
          \and
          A. P\'eer  \at
          Bar Ilan University
          Ramat-Gan-Gan 52900, Israel
	\and
	 S. D. Vergani \at
	GEPI, Observatoire de Paris, PSL University, CNRS, 5 Place Jules Janssen, F-92190 Meudon, France
	\and		
	S. Zane \at
	Mullard Space Science Laboratory, University College London, UK 
          }

\date{Received: date / Accepted: date}

\maketitle

\begin{abstract}
Gamma-ray Bursts (GRBs) are the most powerful transients in the Universe, over--shining for a few seconds all other $\gamma$-ray sky sources. Their emission is produced within narrowly collimated relativistic jets launched after the core--collapse of massive stars or the merger of compact binaries. 
THESEUS will open a new window for the use of GRBs as cosmological tools by securing a statistically significant sample of high-$z$ GRBs, as well as  by providing a large number of GRBs at low--intermediate redshifts extending the current samples to low luminosities. The wide energy band and unprecedented sensitivity of the  Soft X-ray Imager (SXI) and X-Gamma rays Imaging Spectrometer (XGIS) instruments  provide us a new route to unveil the nature of the prompt emission. For the first time, a full characterisation of the prompt emission spectrum from 0.3 keV to 10 MeV with unprecedented large count statistics will be possible revealing the signatures of synchrotron emission. SXI spectra, extending down to 0.3 keV, will constrain the local metal absorption and, for the brightest events, the progenitors' ejecta composition. Investigation of the nature of the internal energy dissipation mechanisms will be obtained through the systematic study with XGIS of the sub-second variability unexplored so far over such a wide energy range.  THESEUS will follow the spectral evolution of the prompt emission down to the soft X--ray band during the early steep decay and through the plateau phase with the unique ability of extending above 10 keV the spectral study of these early afterglow emission phases.

\end{abstract}

\section{Introduction}
\label{intro}

Our picture of Gamma Ray Bursts (GRBs) has evolved in the latest 50 years through giant steps forward enabled mainly by outstanding observational  discoveries from space or ground based facilities. From their discovery in the seventies \cite{Klebesadel1973}, the Burst And Transient Source Experiment (BATSE) on board CGRO revealed, in the 90s, the non--thermal nature of the emission, confirmed their isotropic sky distribution and revealed the existence of the short/long divide \cite{Kouveliotou1993}. The  discovery of the afterglow emission by BeppoSAX \cite{Costa1997} and the measurement of their redshifts    definitely closed the debate about the distance scale to these powerful sources, proving the cosmological origin and implying cataclysmic energies. Soon after, the first direct evidence of the origin of long GRBs from massive star progenitors \cite{Galama1998} was secured. The sample of GRBs with measured redshift progressively increased in number with a major advancement imprinted by the {\it Swift} satellite which among other discoveries, identified the farthest GRBs ever (e.g. GRB 090423 - \cite{Salvaterra2009,Tanvir2009}), detected the first short GRB afterglow and allowed the measurement of their distance scale \cite{Gehrels2005}. {\it Swift} revealed the unexpected temporal structure of the early X--ray afterglow \cite{Tagliaferri2005,Burrows2005} thus opening a new window on the transition between the prompt and the afterglow phase. Since 2008, the {\it Fermi} satellite revealed the emission of GRBs in the 10 keV -- 100 GeV energy range thus probing  the early rise of the afterglow emission at very large energies ($>100$ MeV - \cite{Abdo2009,Ghirlanda2010,Ackermann2015}) and allowing us to systematically study the shape of the prompt emission of long and short GRBs (e.g., \cite{Nava2011,Gruber2014,Calderone2015}). The year 2017 witnessed the opening of the multi--messenger era where GRBs are still among the main characters: the first association of the gravitational wave signal GW 170817 from the merger of two neutron stars with a short GRB 170817 \cite{Abbott2017a,Abbott2017b} confirmed that binary compact mergers can be the progenitors of short GRBs and that in the merger a bright thermal emission, the kilonova, is produced \cite{Pian2017}. Very recently, the ground based Cherenkov telescope MAGIC clearly detected the inverse Compton peak (0.2 -- 1 TeV) of the afterglow emission in the nearby GRB 190114C \cite{Magic2019a,Magic2019b}. 

While the observational picture of GRBs was enriched by both the progressively larger sample of GRBs with measured 
reshifts pinpointed by the Swift satellite and by the study of the emission in the MeV--GeV energy range by the {\it Fermi} satellite, 
the sample of high redshift GRBs still comprises only 6 events at $z>6$ [100]. These GRBs are crucial to probe the 
early Universe. 


THESEUS \cite{Amati2018} combines the rapid slewing capabilities and large field of view of {\it Swift} with the large energy range and effective area of {\it Fermi}. The added values are the high sensitivity of the Soft X--ray Imager (SXI), the wide energy range and large effective area of the X--Gamma rays Imaging Spectrometer (XGIS) and the on-board Infra-Red Telescope (IRT). 
This unique combination of instruments on board THESEUS will allow us to detect high redshift GRBs (and measure their redshifts), and study the prompt and X--ray properties of the full population of detected events.

This paper is organized as follows. In Section~\ref{pop} we describe the population models for both long and short GRBs we developed in order to predict the expected GRB samples detectable by THESEUS. Section~\ref{pr_sp} discusses how the combination of THESEUS X-ray and $\gamma-$ray instruments will provide a new route to study the prompt emission phase and, in particular, to reveal the signature of the synchrotron emission. In Section~\ref{afterglow} we show how THESEUS can monitor the transition from the prompt to the afterglow. In Section~\ref{timing} we present the contribution of THESEUS/XGIS to the study of sub-second variability which can shed light on the nature of the internal energy dissipation mechanism. Conclusions are reported in Section~\ref{conclusions}.

Throughout the paper we adopted the 'concordance' model values for the
cosmological parameters: $h=0.7$, $\Omega_m=0.3$, $\Omega_\Lambda=0.7$,
where $h$ is the dimensionless Hubble constant, $H_0=100 h$ km s$^{-1}$ 
Mpc$^{-1}$.

\section{Populations of GRBs accessible by THESEUS}\label{pop}
Our current knowledge of the population of long and short GRBs is based on  samples of events (thousands of bursts) detected by past and current GRB detectors on board  different satellites. These samples, providing us a statistically rich view of the prompt emission properties of GRBs, are subject to the specific instrumental selection effects, namely the ability to detect, within the detector's energy window, GRBs above a given limiting flux. More specifically, given the prompt emission diversity in terms of temporal variability, overall duration and spectral shape and evolution, GRB samples detected by a specific instrument are biased by its properties such as the (direction-dependent) effective area, the temporal resolution, the instrumental background level, the trigger algorithms, etc. Over the last 20 years, with the discovery of GRB afterglows (started with \cite{Costa1997}), our picture was further enriched by the measurement of redshifts allowing us to access the intrinsic properties of these sources such as the luminosity, spectral peak energy, duration. {\it Swift} has a primary role in this field ensuring fast pinpointing, down to few arcmin precision, of the afterglow. Spectroscopy from ground provides us the redshift of an increasingly larger sample of GRBs, although with an average efficiency of $\sim$30\% for GRBs triggered by {\it Swift}/Burst Alert Telecospe (BAT) \cite{Salvaterra2012}. 

If we aim to estimate the GRB detection rate of an instrument like none before, we cannot rely on these biased samples but rather we should rely on GRB population models which eventually allow us to extend the detection to any combination of source physical parameters, beyond what has been explored so far. 

A population of cosmic sources can be described by the luminosity function (i.e. the number of sources as a function of their luminosity or energy) and the distribution in redshift. For GRBs, a direct measure of these functions is hampered by the many biases that shape current observed samples (e.g. \cite{Pescalli2016}) and the paucity of GRBs with measured redshifts. 
For the study of the science requirements of THESEUS we built a synthetic population of GRBs under   physically motivated assumptions on the shape of the two functions \cite{Ghirlanda2015}. 

The luminosity function of both long and short GRBs is usually parametrised (e.g. \cite{Salvaterra2012,WandermanPiran2010,WandermanPiran2015,Pescalli2015,Ghirlanda2016}) as a double power-law with a faint end slope $\alpha_f$, a bright end slope $\alpha_b$ and a break luminosity $L_{\rm break}$. We extend the luminosity function down to $\sim 1\times 10^{46}$ erg s$^{-1}$ in order to include also low luminosity events like GRB 980425. We assume that the luminosity function is the same at all redshifts. In virtue of the existence of a link between the luminosity and the peak energy, this also ensures that the simulated population includes also very soft GRBs \cite{Pescalli2015} (also known as X—ray flashes -  \cite{Sakamoto2008}). 

The intrinsic redshift distribution depends on the physical conditions that give rise to the GRB event and therefore its shape is different for long and short GRBs.
Long GRBs are now  associated with the core collapse of massive stars by the detection of a type Ib,c supernova associated with almost all long GRB events in the low-redshift Universe where these studies are possible \cite{HjorthBloom2012}. This fact suggests that long GRBs could be good tracer of the star formation. However, both population studies (e.g. \cite{SalvaterraChincarini2007,Salvaterra2012}) and the observed properties of the galaxies hosting the GRB event \cite{Vergani2015,Perley2016,Japelj2016,Vergani2017,Palmerio2019} suggest that their rate increases with redshift more rapidly and peaks at higher redshift than that of stars. These evidences support a scenario in which the GRB event requires a low-metal content in the progenitor star \cite{Heger2003,Yoon2006} and, therefore, their formation is hampered in a metal-rich environment. We model this effect by multiplying the cosmic SFR \cite{MadauDickinson2014} by a factor $(1+z)^\delta$ \cite{SalvaterraChincarini2007,Salvaterra2012,Ghirlanda2015} which increases the GRB--to--Cosmic star formation rate as due to the decrease of the cosmic metallicity.
Short GRBs are one of the outcome of the mergers of compact object (NSNS and possibly NSBH) binaries as shown by the temporal and spatial association of GRB 170817A with the gravitational wave event GW 170817 produced by the merger of two NSs at $\sim$40 Mpc  \cite{Abbott2017a,Abbott2017b}. Therefore, the SGRB redshift  distribution peaks at lower redshifts than the cosmic SFR due to the delay between the formation of the progenitor binary and its merger \cite{Belczynski2006,Salvaterra2008,WandermanPiran2015,Ghirlanda2016}. Due to our poor knowledge of the merger delay time distribution, we assume the intrinsic short GRB redshift distribution  derived in \cite{Ghirlanda2016}.

We assume valid for the population of long and short GRBs the respective prompt emission empirical correlations \cite{Amati2002,Yonetoku2004} between the rest frame $\nu F_{\nu}$ peak energy and the isotropic equivalent luminosity and energy (see \cite{Nava2012,DAvanzo2014} for long and short GRBs respectively).


\begin{figure*}
\hspace{-0.9cm}
\includegraphics[width=1\textwidth]{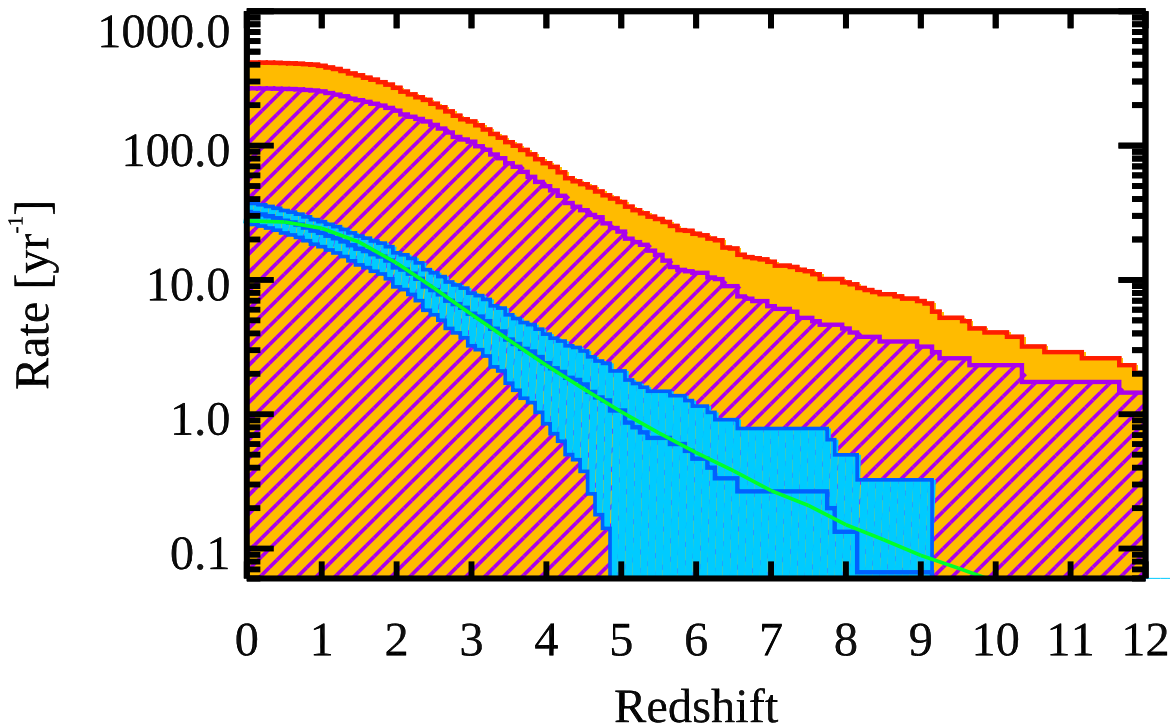}
\caption{Expected detection rate of long GRBs by THESEUS (orange histogram) compared with observed GRBs with redshift measured between 2005 and 2020 (blue line and filled cyan area representing 1$\sigma$ uncertainty). The purple hatched histogram represents the GRBs for which a determination of the redshift by either THESEUS or ground-based facilities is expected. The green curve represents a model fitting the observed distribution on whose basis the THESEUS predictions are made. THESEUS will detect between one and two orders of magnitude more GRBs than {\it Swift} at any redshift, and most notably in the high-redshift regime ($z>6$).}
\label{fig:cumulative}       
\end{figure*}

\begin{figure*}
\includegraphics[width=1\textwidth]{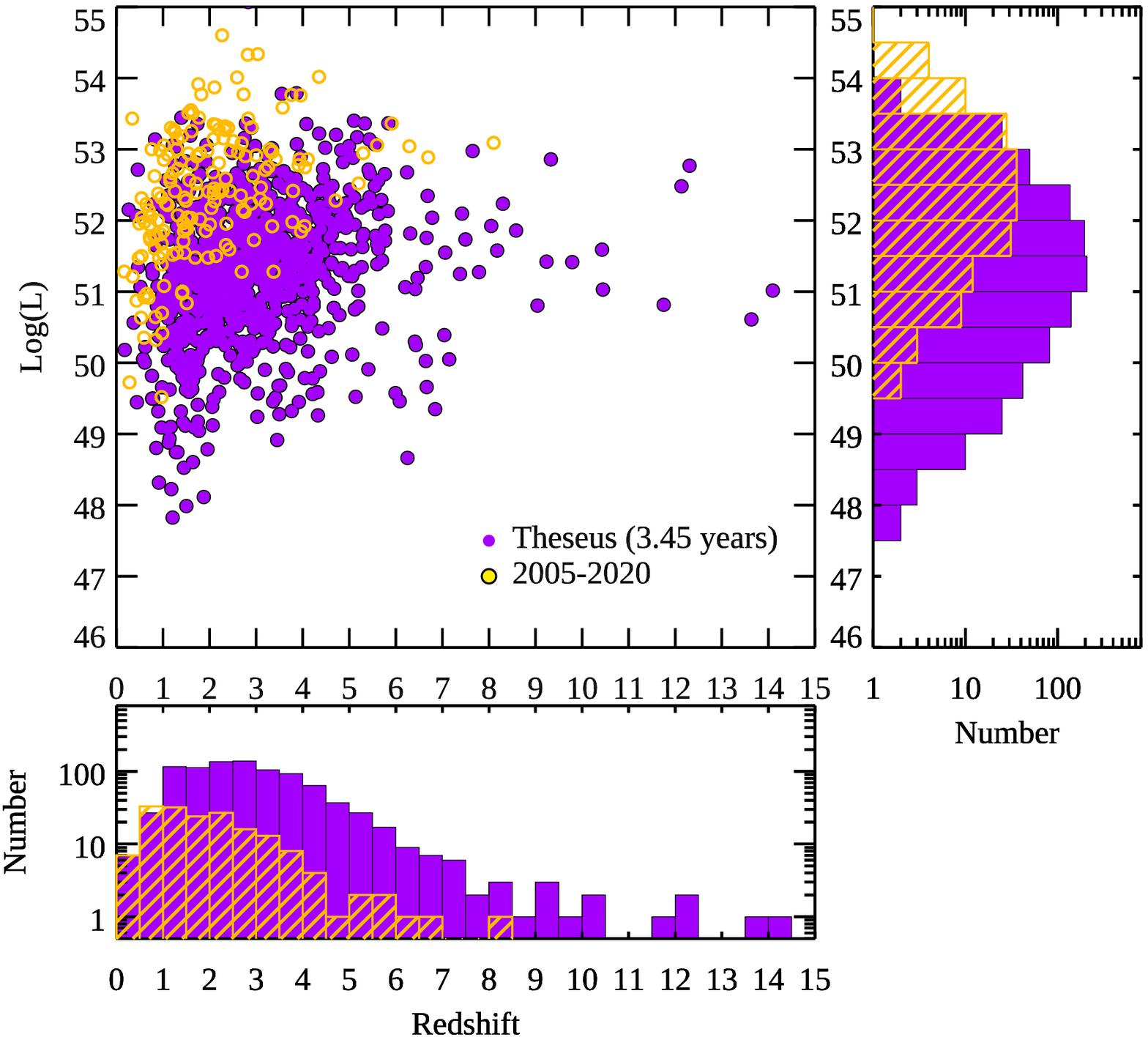}
\caption{Distribution of long GRBs with redshift determination in the peak isotropic luminosity versus redshift plane now (yellow points and hatched histogram) and after the nominal operation life of THESEUS (purple points and full histogram). The right (bottom) panel shows the differential distributions in luminosity (redshift). }
\label{fig:L-z}       
\end{figure*}

\subsection{Monte Carlo simulation of the GRB populations}

We adopt a Monte Carlo approach to simulate the populations of long and short GRBs (LGRB and SGRB hereafter). In both cases, at each simulated event we assign randomly and independently a value of the 
peak energy and a redshift from the assumed distributions. Through the peak energy—isotropic luminosity \cite{Yonetoku2004} and the peak energy—isotropic energy \cite{Amati2002} correlations (accounting also for their scatter) we assign to each simulated burst $L_{\rm iso}$ and $E_{\rm iso}$ respectively. The observer frame fluence and peak flux of the bursts are obtained by assuming a Band \cite{Band1993} spectral shape with  low and high energy spectral slopes randomly extracted based on their observed distributions (e.g. \cite{Nava2012}). This procedure is iterated for $2\times 10^6$ bursts in order to avoid under sampling issues when dealing with relatively steep input functions. The simulated population is then compared with the observed distribution of bursts detected by {\it Swift} and {\it Fermi} accounting for their field of view, mission duration and duty cycle. 

\subsection{Observational constraints and calibration}

The simulations are calibrated by the most updated observed distributions of {\it Swift}
\footnote{https://swift.gsfc.nasa.gov/results/batgrbcat/} and {\it Fermi} Gamma-ray Burst Monitor (GBM)  bursts\footnote{https://heasarc.gsfc.nasa.gov/W3Browse/fermi/fermigbrst.html}. In order to minimise observational biases we consider only relatively bright bursts (selection is made on the peak flux) for which observed samples are complete (\cite{Ghirlanda2015,Ghirlanda2016}). We consider the fluence, peak flux, observed peak energy and duration  distributions of {\it Fermi} GRBs (selected sample contains $\sim 800$ long and $\sim 200$ short GRBs). For the long GRB population we also match the peak flux distribution of {\it Swift} GRBs. For both long and short GRBs we use as constraints the redshift, isotropic equivalent luminosity and energy  distributions of the complete {\it Swift} samples (BAT6 for long – \cite{Salvaterra2012} - and SBAT4 for short – \cite{DAvanzo2014}). The simulated populations are normalized to the detection rate of {\it Fermi} short and long GRBs by accounting for an average duty cycle and the detector field of view. 

Our procedure produces a good fit of all observational constraints when the luminosity function parameters for long GRBs are $\alpha_f\simeq -1.2$ and $\alpha_b\simeq -2.5$ (for the faint and bright end of the function, respectively) with a break luminosity $L_{\rm break}\sim 2\times 10^{52}$ erg s$^{-1}$. The redshift evolution parameter of long GRBs (overimposed on the SFR density evolution) is found to be $\delta=1.7\pm 0.5$ (see \cite{Salvaterra2012,Ghirlanda2015} for details). For short GRBs, we obtain a relatively flat faint end of the luminosity function with slope $-0.5$ and a steeper bright end with slope $-3.4$, the break luminosity being at $\sim 3\times 10^{52}$ erg s$^{-1}$. The intrinsic redshift distribution of short GRBs is consistent with the convolution of the SFR with a delay time distribution $P(\tau)\propto \tau^{-1}$ (see \cite{Ghirlanda2016} for details), where $\tau$ is the delay time between the compact object binary formation and its merger.

\subsection{GRBs detectable by THESEUS}

We use the calibrated synthetic population (of long and short GRBs) together with the knowledge of the THESEUS instrumental properties to estimate the impact of the mission on GRB population studies (see \cite{Mereghetti2021} for further details). The power of THESEUS in detecting and localizing GRBs, and eventually providing a redshift measurement, is shown in Fig.~\ref{fig:cumulative}. The cumulative distribution of GRBs detected by THESEUS through both SXI and/or XGIS (orange histogram) is compared to the distribution of GRBs with measured $z$ between 2005 (the start date of the {\it Swift} mission) and the end of 2020 (blue histogram). The orange distribution is obtained relying on the IRT capabilities for GRBs at $z\gtrsim 5$ and assuming a 50\% successful rate of ground follow up for $z<5$. THESEUS will detect a factor $\sim 10$ more GRBs than {\it Swift}, most notably in the high redshift range ($z>6$). The majority of GRBs detected by THESEUS will have a measured $z$ either on--board and/or from ground.  

The sample of LGRBs detected by THESEUS with $z$ measured (violet symbols) after the nominal 3.45 yr mission is shown in the luminosity--redshift plane of Fig.~\ref{fig:L-z}. Long GRBs detected mostly by {\it Swift} and {\it Fermi} between 2005 and 2020 with measured $z$ and isotropic equivalent luminosity $L_{\rm iso}$ are shown by the open yellow symbols. The two side histograms compare the corresponding $L_{\rm iso}$ and $z$ distributions.  GRBs detected by THESEUS (violet symbols) will outnumber by an order of magnitude the currently known bursts at $z>5$.  Moreover, due to the combined extension of the energy range down to soft X--rays and the combined SXI/XGIS sensitivity, THESEUS will explore at any redshift the luminosity function down to lower luminosities than currently known. 

\begin{figure*}
\hspace{-1cm}
\includegraphics[width=1\textwidth]{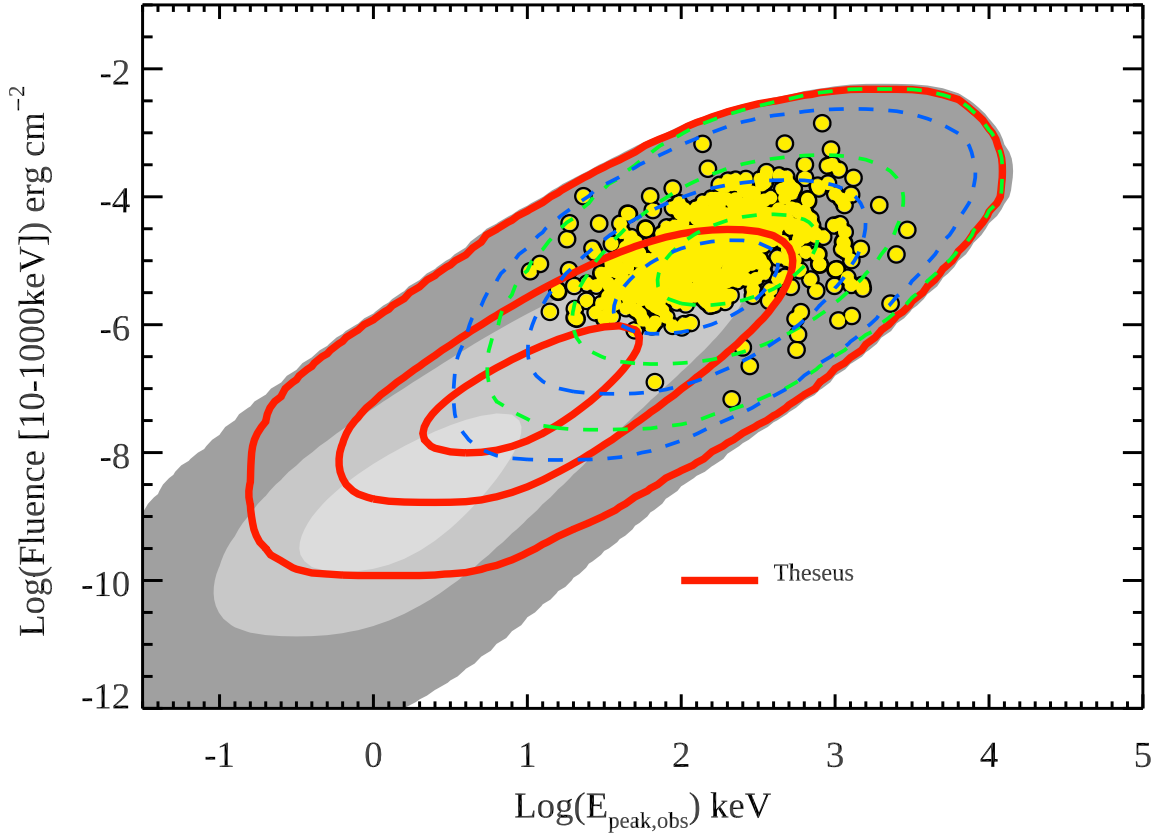}
\caption{The simulated long GRB population in the observer frame showing the correlation between the fluence and the observed peak energy (grey shaded area representing the 1, 2, 3 $\sigma$ contours). Yellow points represent observed long GRB detected by different missions in the last 15 yrs with measured redshift and peak energy. Red contours show the region of the fluence vs observed peak energy plane accessible with THESEUS/SXI. Green and blue dashed contours represent the fraction of the population accessible by {\it Swift} and {\it Fermi} respectively. }
\label{fig:grb_population}       
\end{figure*}

Owing to the softer energy band sampled (0.3-5 keV), THESEUS/SXI will access a softer population of GRBs (red contours in Fig.~\ref{fig:grb_population}) where both soft, low luminosity, events (mostly detected in the low redshift Universe) are present together with high redshift events which would appear softer.  Compared to {\it Swift} and {\it Fermi} (green and blue dashed contours in Fig.~\ref{fig:grb_population}), THESEUS/SXI will contribute to the study of the population of low luminosity GRBs which could be characterized by different physical properties (e.g. opening angle and/or jet bulk velocities). The current knowledge of this population is limited to few objects due to the higher energy range sampled by {\it Swift}/BAT and {\it Fermi}/GBM.  Therefore, the unique capability of THESEUS/SXI to trigger in the X-ray energy band will open a new window in the study of those low-$z$, soft GRBs that are expected to represent the bulk of the total GRB population. On the basis of our population models, we estimate that $\sim$20\% of bursts detected at $z<5$ will have luminosities less than $10^{50}$ erg s$^{-1}$ and a soft spectrum with peak energies $< 50$ keV.  For the first time, we will be able to compare the properties of a statistical sample of soft, low-luminosities GRBs with their cosmological counterpart at intermediate- or high-$z$. This sample will provide unique insights on the nature of soft GRBs/X--ray flashes/X--ray rich events \cite{Sakamoto2008} and of low luminosity GRBs \cite{Liang2007,Pescalli2015,Salafia2016} whose origin is still unclear.  Finally, the ratio between low-luminosity GRBs and supernovae will provide an estimate of the efficiency of massive stars to produce relativistic jets which successfully break out of the stellar envelope \cite{Ghirlanda2013}.

\begin{figure*}
\hskip -1truecm
    \includegraphics[width=1\textwidth]{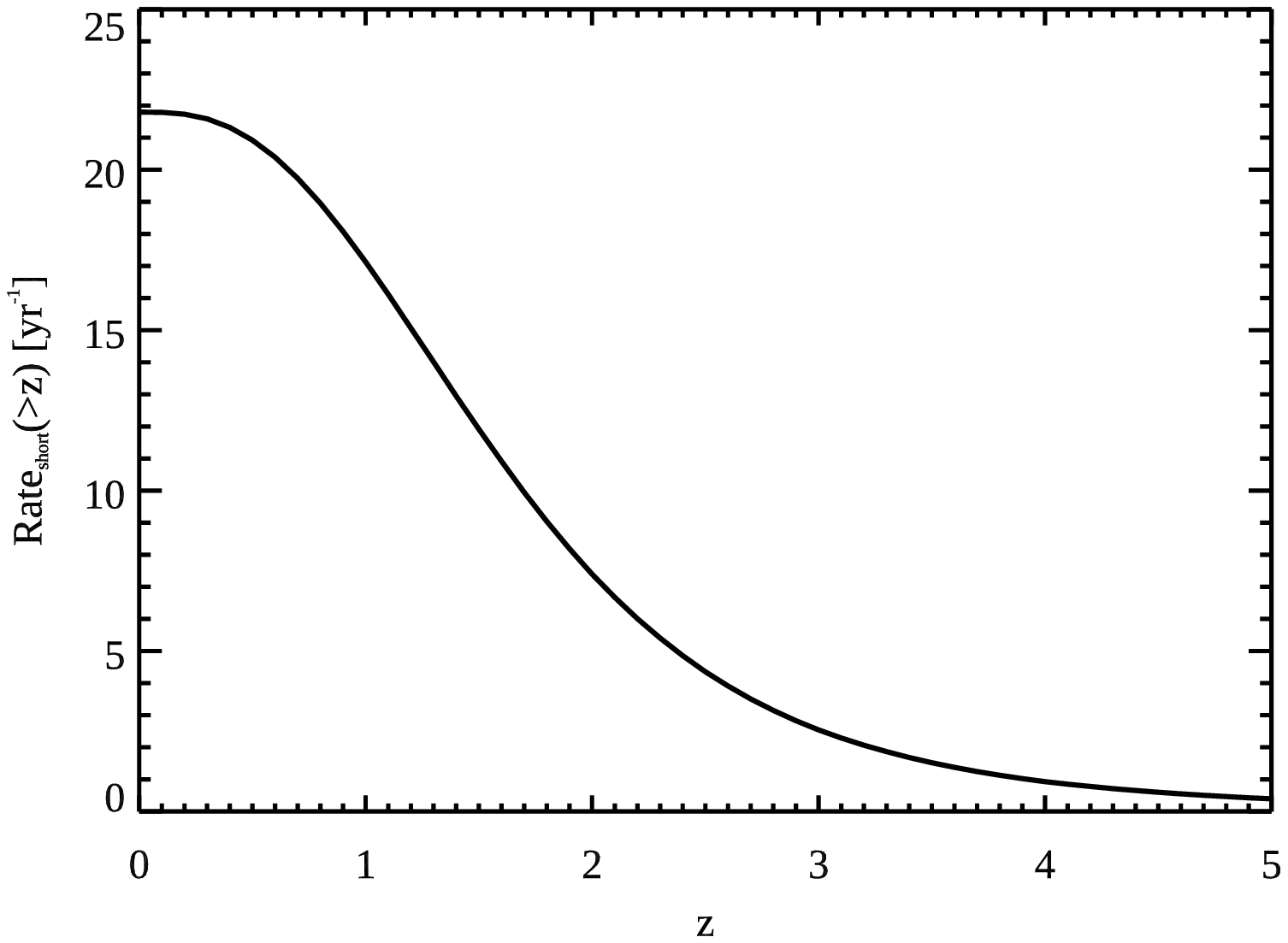}
    \caption{Cumulative redshift distribution of short GRBs detectable with THESEUS/XGIS per year of mission. In principle, short GRBs can be detected even at high redshifts with a rate of $\sim 1$ event per year at $z>4$.}
    \label{fig:short}
\end{figure*}

The cumulative distribution representing the annual detection rate of short GRBs by THESEUS/XGIS is shown in  Fig.~\ref{fig:short}. Note that this is not corrected for mission observation efficiency. 
During its 3.45 years mission, THESEUS will gather a statistically significant sample of short GRBs allowing us to infer the properties of their population. In particular, THESEUS/XGIS should be able to detect short GRBs up to $z\sim 4-5$ so that a more accurate reconstruction of the intrinsic distribution with cosmic times up to the highest redshift would be possible. This will allow us to better constrain the delay time distribution of merger events (e.g. \cite{Ghirlanda2016}).  

\begin{figure*}
\center
\includegraphics[width=1\textwidth]{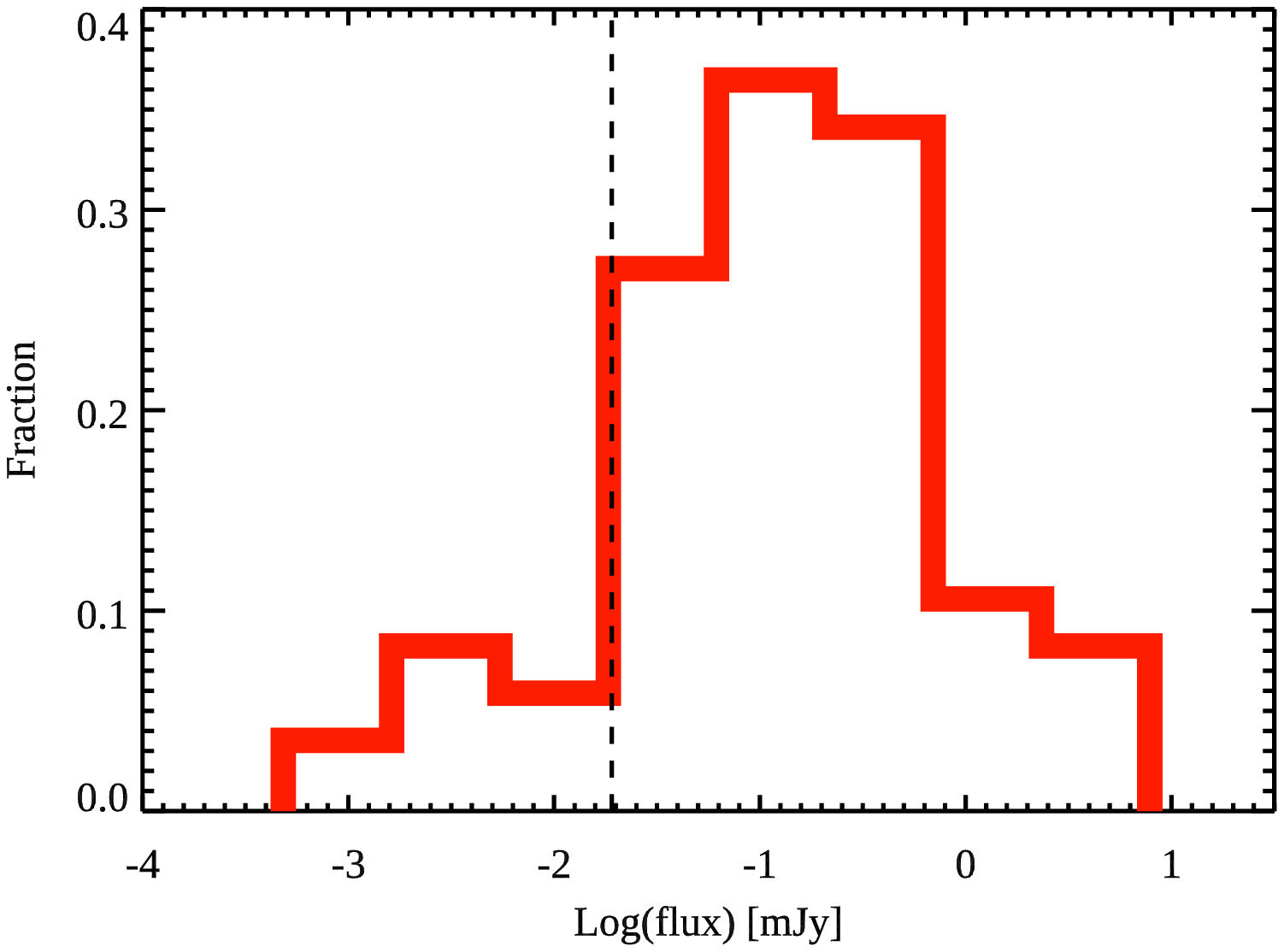}
    \caption{Flux density distribution in the J-band at the peak of the afterglow (considering forward shock emission) for the long GRBs detectable by THESEUS at redshift $z>6$. The vertical dotted line shows the expected limiting magnitude of the IRT with 150 sec exposure. Simulated afterglow flux do not include any correction for possible absorption by gas or dust along the line-of-sight.}
    \label{fig:Jflux}
\end{figure*}

\subsection{Afterglow detection capabilities}

To assess  the THESEUS   requirements of measuring the redshifts of detected GRBs, we added to the simulated population of long GRBs a prediction of the  optical/NIR afterglow flux  based on the emission produced by the deceleration of the fireball in a constant density external medium. We use the model of  \cite{Ryan2020}. The afterglow luminosity depends on the kinetic energy of the jet and on the shock efficiencies in amplifying the magnetic field and accelerating the emitting particles. By following \cite{Ghirlanda2015b}, these parameters are obtained by calibrating the simulated population of bright {\it Swift} GRBs with a complete sample of observed bursts detected by {\it Swift} for which the optical afterglow emission has been sampled between 6 hours and 1 day \cite{Melandri2014}. Furthermore, to each simulated burst we assign a value of the bulk Lorenz factor before the deceleration proportionally to its isotropic prompt emission energy \cite{Ghirlanda2012,Ghirlanda2018} and a jet opening angle randomly extracted from a log--normal probability distribution peaking at 5 degrees and with $\sigma=0.1$ \cite{Ghirlanda2007}. These two parameters allow us to calculate the time of the afterglow deceleration peak and the jet break time which characterize the early and late afterglow light curve, respectively. We consider only the forward external shock emission component. 

Fig.~\ref{fig:Jflux} shows the distribution of J-band fluxes at the peak of the afterglow (approximately reached between 200 and 2000 seconds after the trigger - see also \cite{Ghirlanda2018}) for GRBs at $z>6$ detected by THESEUS/SXI. The vertical dotted line marks the expected limiting flux for a 5$\sigma$  detection in the J band by THESEUS/IRT in 150 s of exposure. Most of high-$z$ GRBs that will trigger THESEUS should be also detected by the IR telescope ensuring the possibility to obtain a precise position of the source in the sky and to determine their photometric/spectroscopic redshifts directly on-board.

\section{The nature of the prompt emission }\label{pr_sp}

The radiative process responsible for the production of the prompt emission of GRBs has not been identified yet (e,g, see \cite{Piran2004,Zhang2020}). The spectral shape of the prompt emission, typically observed in the 10 keV-10 MeV energy range, is inconsistent (e.g., \cite{Preece1998}) with expectations from the synchrotron process \cite{Sari1996,Ghisellini2000}, for typical parameters of the emission region. This triggered the flourishing of a wide range of possible alternatives (e.g. thermal emission from sub-photospheric dissipation, e.g. \cite{Lundman2013}) or modifications of the standard synchrotron scenario (e.g. inverse Compton scattering, peculiar magnetic field configuration  in the emission region, \cite{Rees2005,Asano2009,Peer2017}). 
The lack of our knowledge on the dominant emission processes responsible for shaping the observed prompt emission spectra did not allow us to make a progress in understanding of the physics of GRBs, i.e., wDhere and how the prompt emission is produced in the GRB jets. The identification of the dominant radiative processes would allow us to study in depth the mechanisms of relativistic jet evolution and particles acceleration.   
The extension of the spectral analysis below 10 keV has recently revealed the unexpected presence of a spectral break at low energies, with the bonus of reconciling the overall shape of the prompt spectrum with synchrotron radiation \cite{Oganesyan2017,Oganesyan2018,Oganesyan2019,Ravasio2018,Ravasio2019}.

Besides pointing to the synchrotron process as the mechanism producing the radiation, the detection of this spectral break, interpreted as the signature of the cooling frequency of emitting electrons, opened for the first time the possibility to constrain the physical properties of the emission region (magnetic field intensity, number of emitting particles and their characteristic energy, and localization of the emitting region). However, the identification of this spectral break has been possible only for the small sub-sample of GRBs promptly followed by the {\it Swift}/XRT (e.g. \cite{Oganesyan2018}) or for relatively bright {\it Fermi}/GBM bursts \cite{Ravasio2018,Ravasio2019}. A prototypical example, GRB~180720B at $z=0.65$ (Fig.~\ref{fig:180720B} top left panel) observed by {\it Fermi}/GBM, revealed \cite{Ronchi2020} a steep electron energy distribution (inferred from the slope of the high energy power law spectrum) and a relatively small comoving frame magnetic field, which challenge the acceleration mechanisms, the common understanding of the jet composition, and the standard electron-synchrotron scenario \cite{Ghisellini2020}.

\subsection{Accuracy of spectral parameters}
\begin{figure*}[htbp]
\center
\includegraphics[width=1\textwidth]{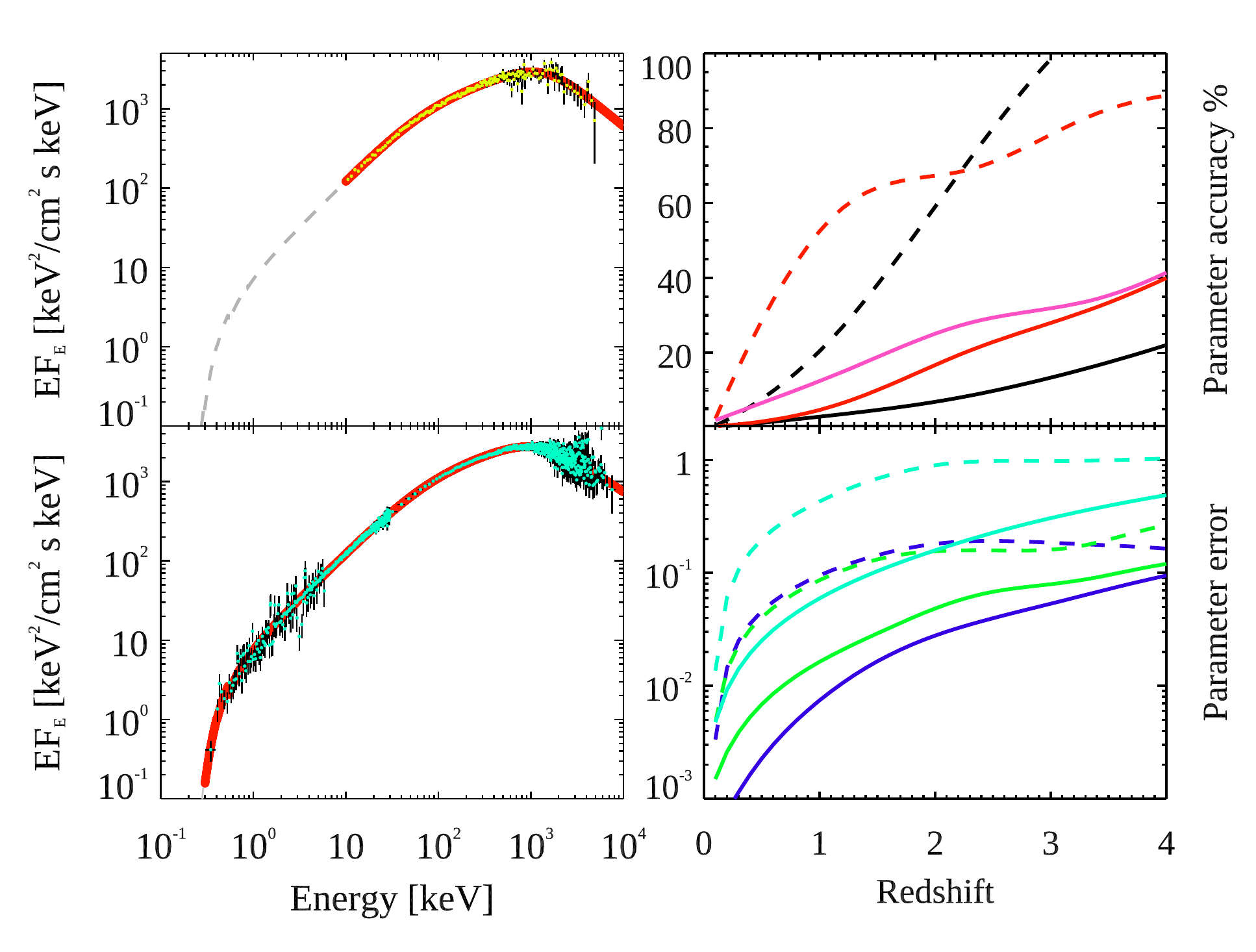}
\caption{Simulated prompt emission spectrum of GRB 180720B \cite{Ronchi2020} as observable by {\it Fermi}/GBM (top left panel) and by THESEUS/SXI+XGIS (bottom left panel). The assumed spectral model is a double break power-law \cite{Ravasio2018} with parameters from \cite{Ronchi2020}. The red line highlight the model in the instrumental energy range. A total absorption $N_{\rm H}=10^{21}$ cm$^{-2}$ is assumed. Right panels: percent precision on model parameters, obtained by simulating GRB 180720B at different redshifts. Top right panel: accuracy on peak energy, break energy and $N_{\rm H}$ (red, black and magenta lines respectively) for THESEUS (solid lines) and {\it Fermi}/GBM (dashed lines). For {\it Fermi}/GBM the $N_{\rm H}$ is undetermined due to the energy threshold of the detectio $>10$ keV. Bottom right panel: absolutes error on the spectral indices of the double break power-law model. The blue, green and cyan lines correspond to the low energy spectral index below the break, to the spectral index of the power-law between the break and the peak and to the spectral index above the peak energy respectively.}
\label{fig:180720B}       
\end{figure*}

THESEUS, by exploiting its larger effective area and wider energy range compared to {\it Fermi}, will systematically observe the full GRB prompt emission spectrum down to $<$1 keV (Fig.~\ref{fig:180720B} bottom left panel) providing the measurement of the low-energy spectral break and its temporal evolution. Using GRB 180720B as a template and moving it to different redshifts (thus accounting for the shift of the peak and break energies and for the decrease of the observed flux), we performed a set of spectral simulations deriving how it would be observed by THESEUS and by {\it Fermi}/GBM. An example of the simulated spectra is shown in the left panels of Fig.~\ref{fig:180720B}. The comparison on the parameter accuracy obtained by fitting the simulated spectra is shown in the right panels of Fig.~\ref{fig:180720B} for THESEUS (solid lines) and {\it Fermi}/GBM (dashed lines). 

The advancement of THESEUS, exploiting its wider energy range (0.3 keV-10 MeV) and larger effective area with respect to {\it Fermi}/GBM, will provide accurate estimates of the key parameters of the prompt emission spectrum and, therefore, of the underlying physical parameters. For instance, at $z\sim$2, corresponding to the peak of the THESEUS-detected population, the break energy will be constrained with an accuracy of 20\% compared to $>$50\% of {\it Fermi} (compare solid and dashed black lines in Fig.~\ref{fig:180720B}, top right panel) with a typical error on the spectral index of the two power-laws below the break of 0.01-0.1 (a factor of 10 smaller than {\it Fermi}). The extension of the spectral window down to 0.3 keV with THESEUS/SXI will provide unique constraints on the Hydrogen equivalent column density ($N_\mathrm{h}$; magenta solid curve in the top right panel of Fig.~\ref{fig:180720B}), thus relieving the degeneracy with low spectral break values.  

\subsection{Identifying the spectral break}

\begin{figure*}[htbp]
    \centering
    \includegraphics[width=\textwidth]{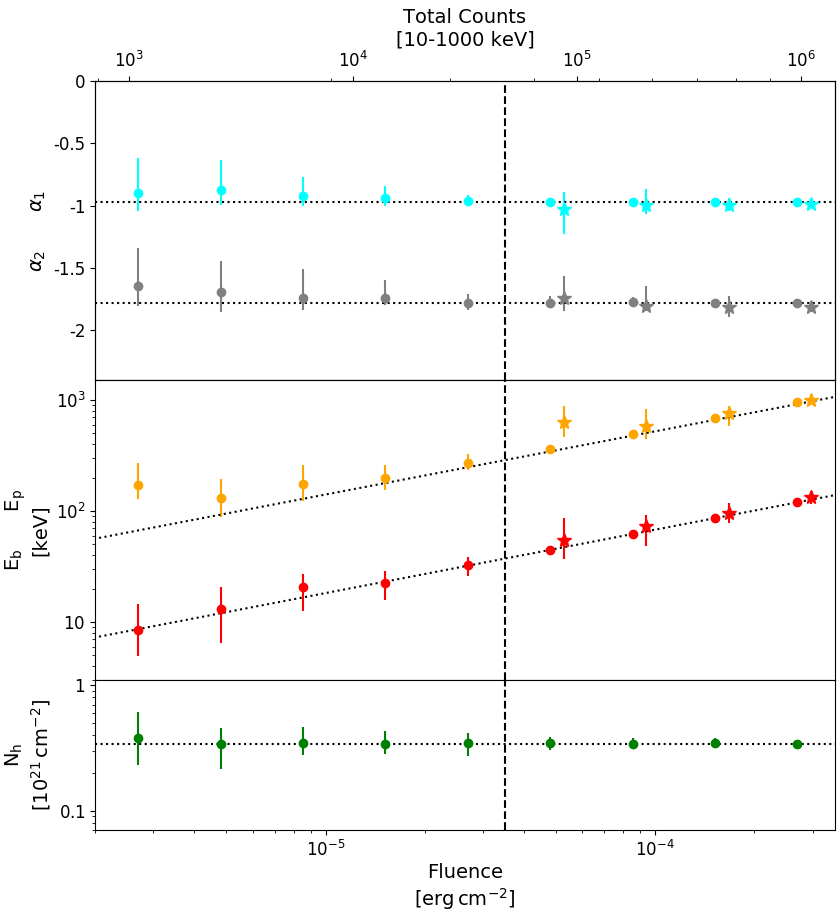}
    \caption{Break detectability as a function of the GRB fluence.   
    Upper panel: values returned by the fit for $\alpha_1$ (cyan) and $\alpha_2$ (grey), the indexes of the power-law below and above the break, respectively; middle-upper panel: values returned by the fit for $E_{\mathrm{break}}$ (red) and $E_{\mathrm{peak}}$ (orange); middle-lower panel: values returned by the fit for $N_\mathrm{H}$ (green). Errors shown are $1\sigma$.
    THESEUS is represented by circles,  {\it Fermi}/GBM by stars (juxtaposed for visualisation purpose).
    The total counts in the spectra are indicated in the upper {\it x}-axis.  Dotted lines indicate input parameters values. The vertical dashed line  represents the minimum fluence for which a spectral break is identifiable with {\it Fermi}/GBM; this fluence is representative of the $\sim 10\%$  brightest GRBs of the Fermi Catalog. THESEUS will be able to reveal the presence of the break in bursts with fluence much smaller than this limit. THESEUS will reveal the presence of the break in the majority of GRBs that it will detect.}
    \label{fig:simul_break}
\end{figure*}


The features of the low energy spectrum of the prompt emission of GRBs hold the key to unveil the nature of the emission process and the energy content of the outflow. To better assess the advancement of THESEUS in characterising the spectral shape below the peak energy, we have performed a set of simulations comparing {\it Fermi}/GBM and THESEUS. We consider GRB~180720B and its parameters as input and we changed the peak energy  $E_{\rm peak}$ and the fluence requiring that the simulated burst moves along the Amati relation \cite{Amati2002}. Based on our current knowledge of the ratio between $E_{\rm peak}$ and the break energy $E_{\rm break}$ \cite{Ravasio2019}, we assumed a constant ratio between these two parameters. 

GRB~180720B is among the GRBs with the largest fluence detected by {\it Fermi} ($\mathcal{F} \sim 2.7\times 10^{-4} \, \mathrm{erg \, cm^{-2}}$). 
For each set of parameters we simulate the corresponding spectrum as it would be observed by {\it Fermi} and by THESEUS and refit with the same input function, namely a double break power--law \cite{Ravasio2018}.
The results of this set of simulations are shown in the panels of Fig.~\ref{fig:simul_break}. We show the spectral parameters obtained by the fit for THESEUS (circles) and {\it Fermi}/GBM (stars) with coloured symbols and the input spectral parameters with dotted lines.  For large fluence values ($\simgt 7\times 10^{-5} \, \mathrm{erg \, cm^{-2}}$) both THESEUS and {\it Fermi}/GBM satisfactorily return the parameters of the double break function used as input. However, {\it Fermi}/GBM can constrain the parameters of this function only down to a fluence $\sim 3\times 10^{-5} \, \mathrm{erg \, cm^{-2}}$. In the sample of GRBs detected by {\it Fermi}/GBM there are only $\sim$ 10\% of bursts with a larger fluence. Below this fluence value, the simulated spectra,  as observed by {\it Fermi}, have not enough counts to constrain the break. We have verified that below this fluence value {\it Fermi} simulated spectra can be best fitted by a Band function, despite the input model was a double break power--law function. 

The analysis of THESEUS--simulated spectra with the double break function gives spectral parameters (circles) which are consistent with the input values (dotted lines).  This holds for spectra with fluence as low as  $\sim 10^{-6} \, \mathrm{erg \, cm^{-2}}$. Therefore THESEUS is expected to find a spectral break, whenever this feature is present, for $>50\%$  of the GRBs that it will trigger. 
Moreover, THESEUS/SXI, by sampling the prompt emission spectrum in the 0.3-5 keV energy range, will constrain the total $N_\mathrm{H}$ with high accuracy (bottom panel in Fig.~\ref{fig:simul_break}). In turn, the acquisition of the prompt emission spectrum by THESEUS down to 0.3 keV with SXI will allow to disentangle the curvature induced at low energies by the spectral break with respect to that produced by the metal absorption. 

Around few tens of keV, where most of the spectral breaks in {\it Swift} bursts have been identified \cite{Oganesyan2017,Oganesyan2018}, it is also possible that a thermal black--body emission component is present \cite{Ghirlanda2003,Ghirlanda2013b,Ryde2011}. The emergence of the dominant thermal emission in the very early phases of the prompt emission \cite{Ghirlanda2003}, its presence as a sub-dominant component for a large fraction of the burst duration or its appearing in the early afterglow phase (e.g. \cite{Nappo2017}) can  provide us unique clues on the outflow dynamics and energy content (e.g. \cite{Daigne2002,Ghirlanda2013b,Peer2015}). THESEUS will systematically reveal this emission component whenever present and clearly disentangle, thanks to the low energy extension by SXI of the spectral range, it from the spectral break \cite{Oganesyan2019}. Moreover, the broad spectral energy range of THESEUS guarantees the detection of the soft X--ray emission arising from the shock break out (e.g. as in GRB 060218 - \cite{Campana2006,Levinson2020}).

\subsection{Metal enriched circumburst medium}

The large luminosity of the prompt GRB emission allows us to probe in absorption the metal content along the line of sight closest to the GRB. This material bears the imprint of the progenitor explosive nucleosynthesis. So far, only in GRB 990705, one of the brightest bursts detected by the BeppoSAX satellite, a transient absorption edge at 3.8 keV in the early phases ($\leq 10$ s) of the prompt emission was discovered \cite{Amati2000}. The absorption edge, detected with the Wide Field Cameras (WFC, 2--26 keV) on board SAX, was interpreted as due to material with a relative iron abundance Fe/Fe$_{\odot}=70\pm19$ at the redshift $z=0.86\pm0.17$. The redshift estimate was later confirmed to agree with that of the host \cite{LeFloch2002}. The origin of this material was attributed to the supernova ejecta thus constraining the explosion time. Absorption edges were not found in other GRBs \cite{Campana2016}, though the search was limited to the early afterglow phase. 

\begin{figure*}
    \centering
    \includegraphics[width=\textwidth]{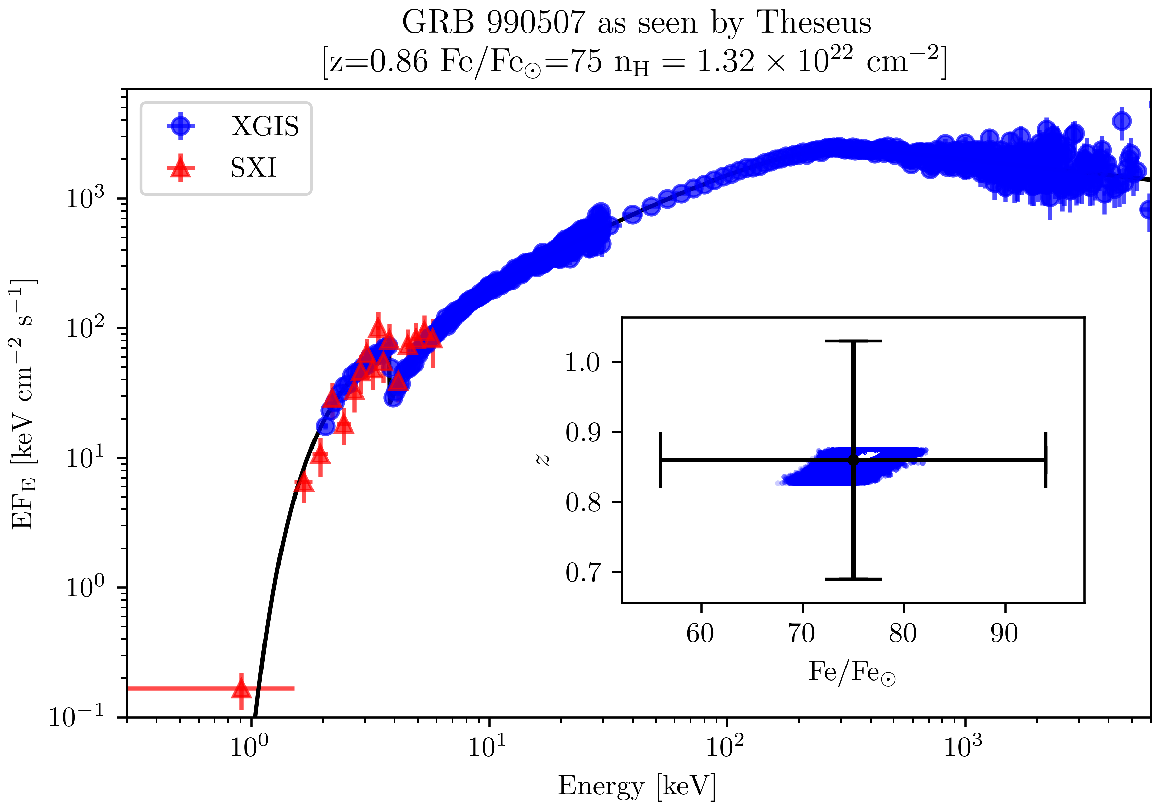}
    \caption{Simulated prompt emission spectrum of GRB 990705 \cite{Amati2000} as observable by THESEUS/SXI (red) and XGIS (blue). The source parameters (in addition to those in the subtitle) as derived in \cite{Amati2000} are assumed for the simulation. Exposure time is 10 seconds. The best fit model (solid black line) shows a prominent absorption edge due to the K-shell Fe transitions at 3.4 keV (observer frame). Insert: posterior 2D constraints (blue shaded region representing the 3$\sigma$ confidence region) on the iron abundance (in solar units) and redshift. The black cross shows the results of  BeppoSAX.}
    \label{fig:990507}
\end{figure*}

Fig.~\ref{fig:990507} shows the spectrum of GRB 990507 as it would be observed by THESEUS. The edge is well detected by SXI (red symbols) and even better sampled by XGIS (blue symbols). The insert of Fig.~\ref{fig:990507} compares, in the $z$--Fe/Fe$_{\odot}$ plane, the 3$\sigma$ constraints obtained from the fit of THESEUS data (blue region) with the constraints obtained from the BeppoSAX/WFC data (black cross).  THESEUS can measure with unprecedented precision the circum--burst metal abundance paving the ground to unprecedented insights into the progenitor’s explosion nucleosynthesis \cite{Fryer2006}. 

\section{Jet structures and early afterglow}\label{afterglow}

\begin{figure*}
    \centering
    \includegraphics[width=\textwidth]{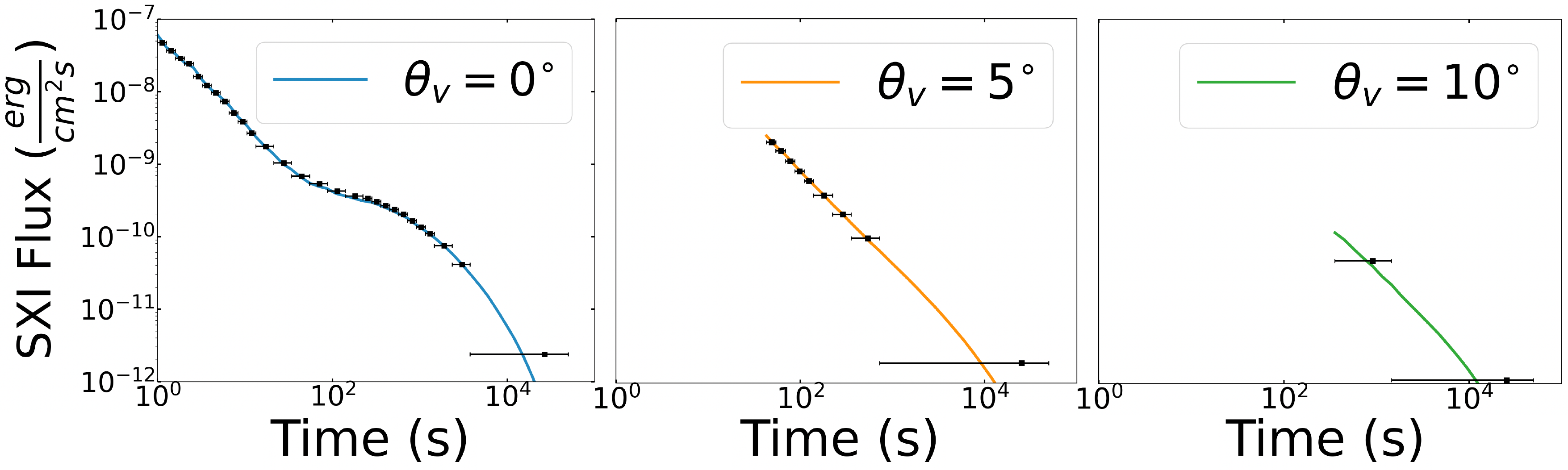}
    \caption{Temporal sampling of a light curve with THESEUS/SXI produced by the high latitude emission of a structured jet, observed at different inclination angles with respect to the symmetry axis of the jet. Each light curve is rescaled at redshift $z=0.5$.}
    \label{fig:stru_jet}
\end{figure*}

THESEUS, combining SXI and XGIS, will systematically monitor the spectral evolution of the prompt gamma-ray emission towards softer energies and will cover the transition from prompt to afterglow. 

With the launch of {\it Swift} the early afterglow emission in the X--ray (0.3-10 keV) energy band has been systematically monitored and revealed a more complex behaviour of the single temporal power-law decay of the afterglows observed by BeppoSAX. A large fraction of X-ray afterglows starts with a steep decay phase $F_X \propto t^{-3}$, followed by a long (up to $10^{4}$ s) flat segment, so called plateau phase $F_X \propto t^{-0.5}$. On timescales of few hours the light curve eventually follows the typical afterglow (e.g., \cite{Zhang2006}). 

The steep decay phase can be interpreted as due to the switching off of the prompt emission. Specifically, the temporal delay/energy softening of photons emitted at large angles with respect to the line of sight (so called high latitude emission, \cite{Fenimore1996,KumarPanaitescu2000}) can reproduce the steep flux decay. A possible further effect which could have a relevant role in shaping the observed steep decay phase is the spectral evolution of the prompt emission consisting in the overall softening of the spectrum \cite{Ronchini2020}. Such a softening of the spectrum can be probed at present only through the X--ray 0.3-10 keV energy band sampled by {\it Swift}/XRT.  

The most discussed model for the  X-ray plateau emission is the fast spinning (ms period) and highly magnetized (B $\sim 10^{15}$ G) newly born neutron star as the central GRB engine (e.g., \cite{Dai1998}). Its spin-down power provides an additional energy injection into the decelerating blast wave producing the plateau emission (e.g., \cite{Zhang2006}). However, this model faces several observational and theoretical issues (e.g., \cite{Fan2006}). An alternative scenario invokes the high latitude emission from a structured jet \cite{Oganesyan2020,Panaitescu2020} to explain the plateau phase. In  the structured jet  model, the emission received from the jet wings (outside of the jet core), is less beamed due to the decrease of the bulk Lorentz factor, which produces the observed flattening of the light curves. This model is capable of producing different X-ray light curves, including those with a post-plateau sudden drop. Jet structure is a natural consequence of the propagation of the jet in the GRB ejecta (e.g., \cite{Aloy2000,Lipunov2001,Rossi2002} - either in the single massive star and in the compact merger progenitor scenario). Among the most convincing observational evidences, the afterglow modelling of GRB~170817 requires a Gaussian jet structure  consisting in an energetic core, of aperture  $\sim$3.5 degrees, surrounded by less energetic and slow wings \cite{Ghirlanda2019} observed at $\sim$20 degrees off axis. If the currently observed plateaus are caused by the out-core emission of the structured jets, then we could detect the bright X-ray emission also if we observe the GRB jet off-axis \cite{Ascenzi2020}. 

In Fig.~\ref{fig:stru_jet} we show how THESEUS/SXI can sample the high-latitude emission from a structured jet, seen at different viewing angles. The sampling is obtained considering the flux-sensitivity curve of SXI and the duration of each time bin is computed such that the average flux is above the minimum detectable flux. For the computation of the theoretical light curve we followed the approach of \cite{Ascenzi2020}, adopting a jet structure with the following properties:
$$
\Gamma(\vartheta)=(\Gamma_0-1)\frac{1}{1+(\vartheta/\vartheta_c)^s}+1
$$
$$
\epsilon(\vartheta)=\frac{1}{1+(\vartheta/\vartheta_c)^s}
$$
where $\Gamma(\vartheta)$  and $\epsilon(\vartheta)$ the structures of the bulk Lorentz factor and the emissivity, respectively, with $\vartheta_c=2^{\circ}$, $\Gamma_0=200$ and s=2.5. We also assumed an emission radius $R_0=5\times 10^{15}$ cm. The soft X-ray peak luminosity produced by the burst is assumed to be 0.001$\times L_{\rm iso}\sim 10^{49}$ erg/s, where $L_{\rm iso}$ is a representative value of the isotropic luminosity \cite{Davanzo2012,DAvanzo2014}. The light curve then is renormalized taking a redshift $z=0.5$. 

THESEUS/SXI can monitor the whole steep decay-plateau phase in the on-axis case (left panel in Fig.\ref{fig:stru_jet}), but it is also able to detect the off-axis emission up to a viewing angle few times larger than the jet core. Therefore, THESEUS will probe the off-core GRB population, whose light curves will peak in the soft X-ray band but will likely be orphan of a detectable $\gamma$-ray emission. Moreover, the systematic study of the on- and off-axis emission on a wide population of GRBs will allow to constrain the jet structure and test its universality. This in turn, will aid population studies in constraining the intrinsic local GRB rate. 

\begin{figure*}
    \centering
    \includegraphics[width=\textwidth]{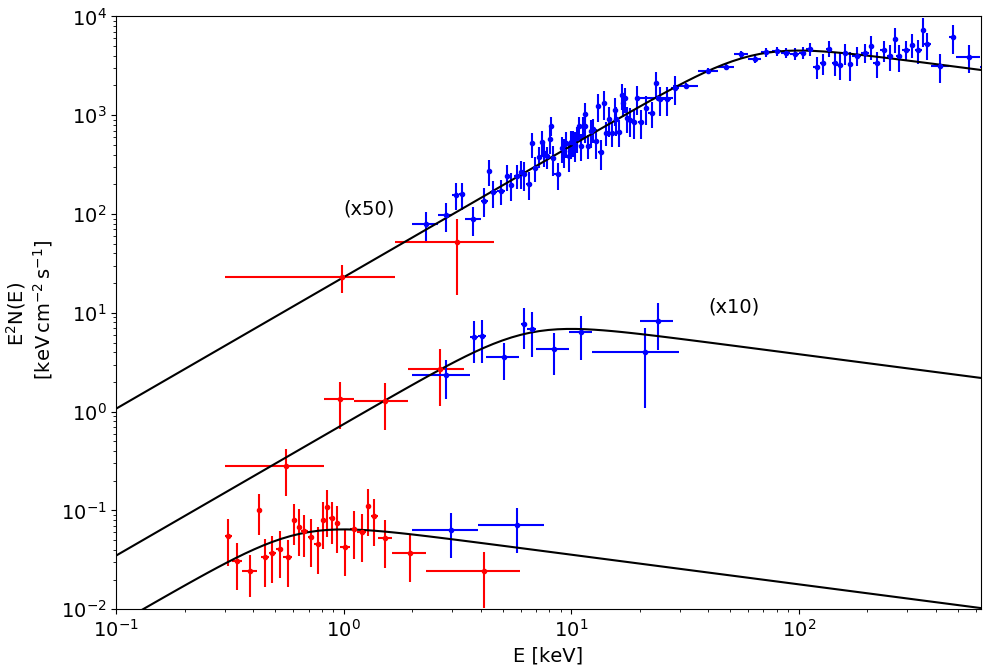}
    \caption{Simulated  spectra during the steep-decay phase (upper spectrum - here multiplied by a factor 50 for the ease of presentation), at the  beginning (middle spectrum - multiplied by a factor 10) and at the end (lower spectrum) of the plateau phase (left panel of Fig. \ref{fig:stru_jet}). The peak energies at 100, 10 and 1 keV are well sampled by THESEUS. SXI and XGIS data are represented by red and blue points respectively.}
    \label{fig:plateau_spec}
\end{figure*}
\begin{table*}[]
    \centering
    \begin{tabular}{c|c|c|c|c|c|c}
    Time  &  $\alpha_\mathrm{Input}$ &   $\beta_\mathrm{Input}$ &   $E_\mathrm{p, Input}$ &$\alpha$ & $\beta$ & $E_\mathrm{p}$ \\
    
    [s]&&&[keV]&&&[keV] \\ \hline
     10 & -0.67 & -2.3 & 100  & $-0.74\pm0.04$ &  $-2.27\pm0.08$& $120^{+10}_{-8}$\\
     100 & -0.67 & -2.3 & 10  &$-0.72^{+0.27}_{-0.25}$& $<-2.20$& $7.8^{+3.5}_{-1.6}$ \\
     1000 & -0.67 & -2.3 & 1  &$-1.12^{+0.50}_{-0.28}$&  $<-2.3$ & $1.87^{+0.58}_{-0.47}$\\ 
     \hline
    \end{tabular}
    \caption{Input parameters and refit results of simulated spectra of Fig. \ref{fig:plateau_spec}. Errors are calculated at 1$\sigma$ confidence. }
    \label{tab:spe_ana}
\end{table*}

The plateau phase has been monitored through the X--ray 0.3-10 keV energy range by {\it Swift}/XRT. Due to the limited bandpass of XRT, in most cases the spectrum during the plateau phase is fitted with only a single power-law function. As a proof of concept we simulate (Fig. \ref{fig:plateau_spec}) the spectrum during the steep-decay ($t_1\sim$10 s), at the beginning ($t_2\sim$100 s) and at the end ($t_3\sim 10^3$ s) of the plateau phase of the light curve (assuming $z\sim0.5$) reported in the left panel of Fig.~\ref{fig:stru_jet}. The integration time for each spectrum is 5 s, 50 s and 500 s, respectively. This simulation assumes the source is on axis in one XGIS unit. We assume that at each time the spectrum is described by a smoothly broken power law function, with low- and high-energy photon indexes $\alpha=-0.67$ and $\beta=-2.3$, respectively. An evolution with time has been assumed for the energy peak $E_{\mathrm{peak}}\sim t^{-1}$. Specifically, we adopt $E_{\mathrm{peak}}(t_1)=100$ keV, $E_{\mathrm{peak}}(t_2)=10$ keV and $E_{\mathrm{peak}}(t_3)=1$ keV. 
The spectral analysis of the simulated spectra returns well constrained parameters (Tab.~\ref{tab:spe_ana}), allowing the study of the spectral evolution at those epochs. The spectral coverage and the large effective area of SXI and XGIS allow to monitor the spectral evolution in the transition from the prompt emission to the afterglow phase, in particular during both the steep-decay and plateau phases, providing clues on their origin.


\section{Temporal properties of GRBs}\label{timing}

\begin{figure*}
    \centering
    \includegraphics[width=\textwidth]{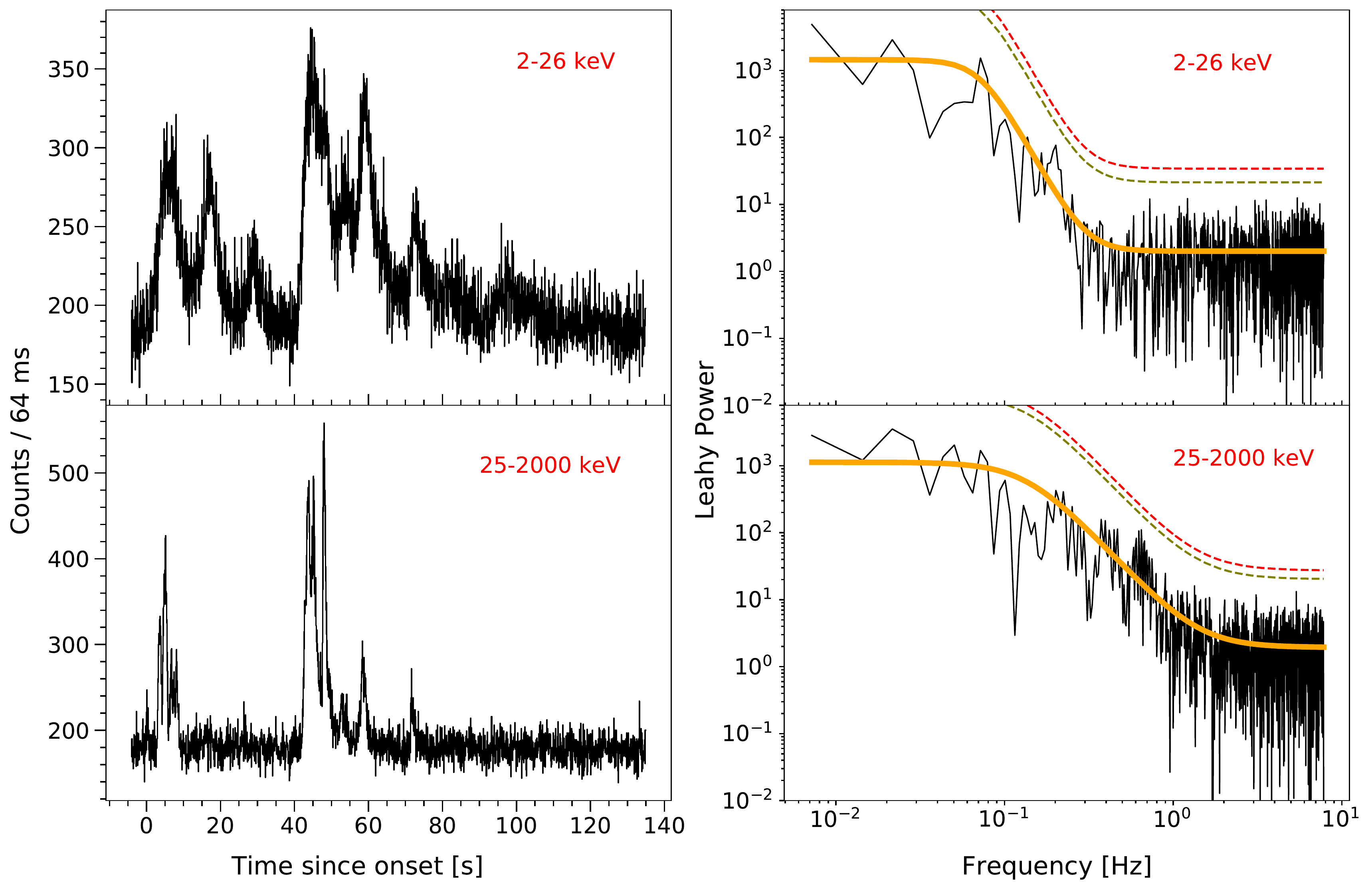}
    \caption{Left: X-ray (top) and gamma-ray (bottom) time profiles expected for a GRB like GRB 990510 that would be observed on axis in one XGIS unit. The integration time is 64 ms and we assumed a constant background. Right: corresponding power density spectra of the time profile alongside. The thick orange line is the best-fitting broken power law model as defined in \cite{Guidorzi2016}; the dashed olive and red lines are the 2 and 3 sigma thresholds for periodic pulsations (see \cite{Guidorzi2016} for details).}
    \label{fig:timing}
\end{figure*}
The variety of GRB light curves along with early-time X-ray flares have poorly been understood, thus remaining mostly undeciphered. On the one side, some metrics that quantify the degree of variability are found to correlate with luminosity with considerable scatter, though \cite{Reichart2001,Rizzuto2007}. Also, power density spectra, both average and individual, are suggestive of turbulence \cite{Beloborodov1998,Guidorzi2012} and correlate with $E_{\rm peak}$ \cite{Dichiara2016}. On the other side, a characterisation as a stochastic process is still in its infancy, mostly hampered by the highly non-stationary, short-lived nature. In this context, it was found that the common waiting time distribution of GRB gamma-ray pulses and X-ray flares can be interpreted as the result of a unique time-dependent Poisson process, providing clues on the way the inner engine works \cite{Guidorzi2015}. Attempts to find evidence for self-organised criticality processes, that are invoked in the cases of solar and stellar flares, pulsar glitches, or even in the asteroid belt and in Saturn rings in the Solar System (see \cite{Aschwanden2018} for a review) have provided some clues from the study of energy, duration, and waiting time distributions of GRB X-ray flares \cite{WangDai2013}. Yet, a much larger data set for a statistically robust assessment is required. Other attempts have also been made to find evidence for the presence of deterministic signal as opposed to a pure random one \cite{Greco2011}. In this respect, the possible presence of periodic signal hidden in GRB light curves has found no compelling evidence \cite{DeLuca2010,Cenko2010,Dichiara2013,Guidorzi2016}. 

The delay in the arrival times between high and low energy photons from cosmic sources can be used to test the violation of the Lorentz invariance (LIV), predicted by some quantum gravity theories, and to constrain its characteristic energy scale EQG that is of the order of the Planck energy. GRBs and blazars are ideal for this purpose thanks to their broad spectral energy distribution and cosmological distances (\cite{Bernardini2017}; see also \cite{Bolmont2008,Ellis2008}). THESEUS spectral coverage, large effective area and relatively high temporal resolution will contribute to test LIV possibly through statistical studies that exploit the samples of long and short GRBs with measured redshift collected by THESEUS. 

Simulating the time profiles of a multi-peaked GRB in the different energy channels of a broad band like 2-2000 keV of THESEUS/XGIS is rather challenging, not only for the as-yet poorly understood nature of the stochastic process that rules them, but also for the paucity of full prompt X-ray profiles available from previous experiments. Among the different approaches that one may opt for, we can identify two major alternative ones: (i) assuming a given sequence of discrete emission episodes, ruled by some random process and described by some distributions (e.g., in the released energy, spectral hardness), where each single episode marks the dissipation of energy and can be derived either theoretically or be just empirically represented with a FRED pulse e.g., \cite{Pescalli2018}; (ii) using an observed GRB as a template, making sure to reproduce the variability properties and their dependence on the energy passband, without introducing artificial variance.

In this context, we here opted for (ii) and looked for a multi-peaked GRB that had been observed in a comparably broadband from past experiments, and found a suitable example in GRB 990510. Its prompt X-ray emission was observed with one of the BeppoSAX Wide Field Cameras (WFCs) in the 2-26 keV energy band \cite{Pian2001,Amati2002,Frontera2012} and was also observed in the 25-2000 keV with the Compton Gamma-Ray Observatory (CGRO) of BATSE. We therefore took its background subtracted time profiles in seven contiguous energy channels, three from the BeppoSAX/WFC (2-5, 5-10, and 5-26 keV), 
 and the remaining four from CGRO/BATSE (25-55, 55-110, 110-320, and 320-2000 keV).
The integration time was 1 s (64 ms) for the WFC (BATSE). Since an observed time profile is inevitably affected by count statistics noise, it cannot be used straightaway as the expected curve of a random realisation, since it would add artificial variance in excess of the expected Poissonian one. We therefore reduced the statistical noise by using a filter \cite{Politsch2020} that optimally works with smoothly varying time profiles that occasionally exhibit spikes and sharp features, such as most GRBs do (see Guidorzi et al. in prep for a detailed description). The filtered WFC profiles were re-sampled with 64-ms resolution by means of C-splines.

We then used the best-fit model obtained by \cite{Amati2002} for the broadband time-average spectrum of GRB 990510 as measured with BeppoSAX WFC and Gamma-Ray Burst Monitor (GRBM), and adopting the 2 -- 2000 keV fluence of $(1.580\pm 0.007)\times 10^{-5}$~erg~cm$^{-2}$ \cite{Frontera2012}, we calculated the fluences corresponding to the energy channels of each time profile. The filtered time profiles were then renormalised so as to yield the corresponding total net counts expected for a GRB like GRB 990510 that would be seen on axis to one THESEUS/XGIS unit, using the appropriate response function (v7). Finally, each renormalised profile was added to the corresponding constant background counts and added the corresponding Poisson noise.

The complex broadband spectral variability exhibited by the GRB is automatically accounted for by treating each individual energy passband separately. We finally summed up the X-ray profile (2-26 keV; hereafter X band) and the gamma-ray one (25-2000 keV; hereafter G band) and calculated the corresponding power density spectra (PDS) from $-4.0$ to $135$~s adopting the Leahy normalisation \cite{Leahy1983}. Both PDS were fitted using the Bayesian MCMC procedure that were used for individual GRB light curves \cite{Guidorzi2016}. In both cases the PDS was found to be best fit with a broken power-law model (Fig.~\ref{fig:timing}). In particular, the break frequency $f_b$ was found to be $7.2^{+1.7}_{-1.4}\times 10^{-2}$~Hz ($0.14\pm0.03$~Hz) for the X (G) band. The corresponding dominant timescale is therefore $\tau_b = 1/(2\pi\,f_b) = 2.2_{-0.4}^{+0.5}$~s $(1.2_{-0.3}^{+0.2}$~s) for the X (G) band.

This confirms and quantifies in an objective way the property that lower energy bands miss temporal power at shorter timescales compared with the harder bands. Only by systematically measuring GRB profiles in a broad band like this will allow for a characterisation of the X-ray prompt emission variability, that will help constrain the dissipation mechanism. 
In addition, the power-law index is also found to be significantly different: $\alpha=4.6\pm 0.6$ (X) vs. $\alpha=2.7\pm0.2$. While the index of the gamma-ray profile is commonly found (e.g., \cite{Guidorzi2016,Dichiara2016}), the index found in the as-yet poorly unexplored X-ray band stands out. One should note that the possible presence of subsecond variance in the X-ray profile was inevitably suppressed since we started with 1s-binned BeppoSAX/WFC time profiles. However, the goal here is just to illustrate the potential of XGIS in this context. Being able to characterise the broad band variability of GRB profiles routinely, as THESEUS/XGIS would ensure, holds great promise to understand the dissipation process that drives that X-ray and gamma-ray emission of GRBs.

\section{Conclusions}\label{conclusions}

THESEUS will open a new window for the detection and study of GRBs over cosmic times. Through physically motivated population models calibrated with available prompt emission data gathered by {\it Swift} and {\it Fermi}, we evaluate the properties and redshift distributions of both long and short GRBs detectable by THESEUS. We showed that THESEUS will enlarge the sample of high-$z$ long GRBs by at least one order of magnitude with respect to what we have now. Moreover, a sizeable sample of low-luminosity GRBs at low- and intermediate-$z$, will allow us to explore the possible existence of a sub--population of soft X--ray flashes  and to study the faint end of the luminosity function which can hold the key of the structure of GRBs jets \cite{Pescalli2015,Salafia2016}.

Owing to the wide energy band and high sensitivitiy of the SXI/XGIS instruments, THESEUS represents the best facility to unveil the nature of the prompt emission. The study of the prompt emission spectrum of the GRBs detected by XGIS will reveal the characteristic signature of the low energy spectrum where synchrotron emission features (such as the particle cooling frequency)  or the photospheric thermal peak lies. Moreover, the extension down to 0.3 keV, provided by THESEUS/SXI, can constrain the local metal absorption and, for the brightest events, progenitors' ejecta composition encoded in metal absorption features. 

Thanks to the timing capabilities of THESEUS/XGIS, the sub-second variability of the prompt emission will be studied over an unexplored wide energy range possibly revealing the nature of the internal energy dissipation mechanisms. Finally, the combination of SXI and XGIS during the steep-flat flux decay following the end of the prompt emission can reveal the structure of the jet and uniquely probe the spectral transition from the prompt to the early afterglow phase.   

In conclusions, THESEUS holds the key to advance in the study of GRBs and to open the high redshift Universe study through the most powerful transients. 


\begin{acknowledgements}
We acknowledge support by ASI-INAF agreement n. 2018-29-HH.0
JO acknowledges partial support from STFC. GG acknowledges support by Premiale project FIGARO 1.05.06.13 and INAF-PRIN 1.05.01.88.06. PDA and GG acknowledge support from PRIN-MIUR 2017 (grant 20179ZF5KS); PDA acknowledge funding from the Italian Space Agency, contract ASI/INAF n. I/004/11/4. We are grateful to Matteo Guainazzi for the fruitful discussions, enlighting comments and suggestions that helped shaping and improving this work. We thank the THESEUS ESA Study Team  and the Science Coordination Office for the support.

\end{acknowledgements}

%
%

\bibliographystyle{spmpsci}      
\bibliography{WG4paper}   

%
%

\end{document}